\documentclass[aps,reprint,12pt]{revtex4}

\usepackage{commath}
\usepackage{amsmath}
\usepackage{graphicx}
\usepackage{hyperref}
\usepackage{txfonts}

\draft 

\begin{document}


\title{Possible self-amplification channel for surface plasma waves}


\author{Hai-Yao Deng$^{1,2}$}
\email{haiyao.deng@gmail.com}
\author{Katsunori Wakabayashi$^{1,3}$}
\email{waka@kwansei.ac.jp}
\author{Chi-Hang Lam$^4$}
\email{c.h.lam@polyu.edu.hk}
\affiliation{$^1$Department of Nanotechnology for Sustainable Energy,
School of Science and Technology, Kwansei Gakuin University, Gakuen 2-1,
Sanda 669-1337, Japan}
\affiliation{$^2$Department of Physics and Astronomy, University of Exeter, Stoker Road EX4 4QL Exeter, United Kingdom}
\affiliation{$^3$National Institute for Materials Science (NIMS), Namiki 1-1, Tsukuba 305-0044, Japan}
\affiliation{$^4$Department of Applied Physics, The Hong Kong
Polytechnic University, Hung Hum, Hong Kong}  


\begin{abstract} 
Surface plasma waves (SPWs) have been extensively studied in the past two decades with a promise for many applications. However, the effort has so far been met with limited success. It is widely believed that a major caveat lies with the energy losses experienced by SPWs during their propagation. To compensate for the losses, amplifiers have been designed, which are all extrinsic and need an external agent to supply the energy. Here we theoretically show that there exists an intrinsic amplification channel for SPWs in the collision-less limit. We pin down the origin of this channel and analytically calculate the amplification rate. Our finding unveils a hitherto unchartered yet fundamental property of SPWs and may bear far-reaching practical consequences.
\end{abstract}


\maketitle 

Collective electronic oscillations on the surface of metals,
dubbed surface plasma waves (SPWs)~\cite{ritchie1957,ferrell1958,raether1988}, have emerged as a pivotal player in \textit{nano}scopic manipulation of light~\cite{benno1988,zayats2005,maier2007,ozbay2006,mark2007}. The functionality of many prototypical nanophotonic devices critically relies on the distance SPWs can travel before they are damped out due to energy losses~\cite{ozbay2006,mark2007,barnes2003,barnes2006,ebbesen2008,martino2012}. Several loss channels exist, most of which can be efficiently but not totally obstructed under appropriate circumstances.  Amplifiers have been contrived to compensate for the losses~\cite{bergman2003,seidel2005,leon2008,leon2010,yu2011,pierre2012,fedyanin2012,cohen2013,aryal2015}, which, however, are all extrinsic and require external agents such as a dipolar gain medium to supply the energy. 

In the present work we report a linear instability analysis within Drude-Boltzmann's theory that reveals an intrinsic amplification channel for SPWs. We find that the eigenfrequencies of SPWs in the collisionless limit are complex with a positive imaginary part $\gamma$, which implies that SPWs are intrinsically unstable. We identify ballistic electrical currents as the origin of the instability. These currents allow a net amount of energy to be drawn from the electrons to the waves. As such, SPWs can self-amplify at a rate $\gamma$ if all loss channels are closed. This observation fundamentally changes our perception of SPWs and other surface waves. 

It might sound counter-intuitive that a wave governed by time reversal symmetric dynamics can exhibit a sense of direction in time. When Laudau predicted that plasma waves must damp out even in the collisionless limit~\cite{landau1946}, a great controversy was stirred which had persisted until the prediction was actually observed in experiments~\cite{wong1964}. Nowadays Landau damping forms a core concept in plasma physics and has borne impact on other subjects~\cite{bingham1997,natu2013}. Our prediction that SPWs can self-amplify is analogous, the difference being that the direction of time is reversed: collisionless SPWs spontaneously amplify rather than damp out. While Landau damping is caused by slowly moving electrons that strip energy from the wave, the predicted SPW self-amplification is attributable to ballistically moving electrons that impart energy to the wave.

\begin{figure*}
\begin{center}
\includegraphics[width=0.95\textwidth]{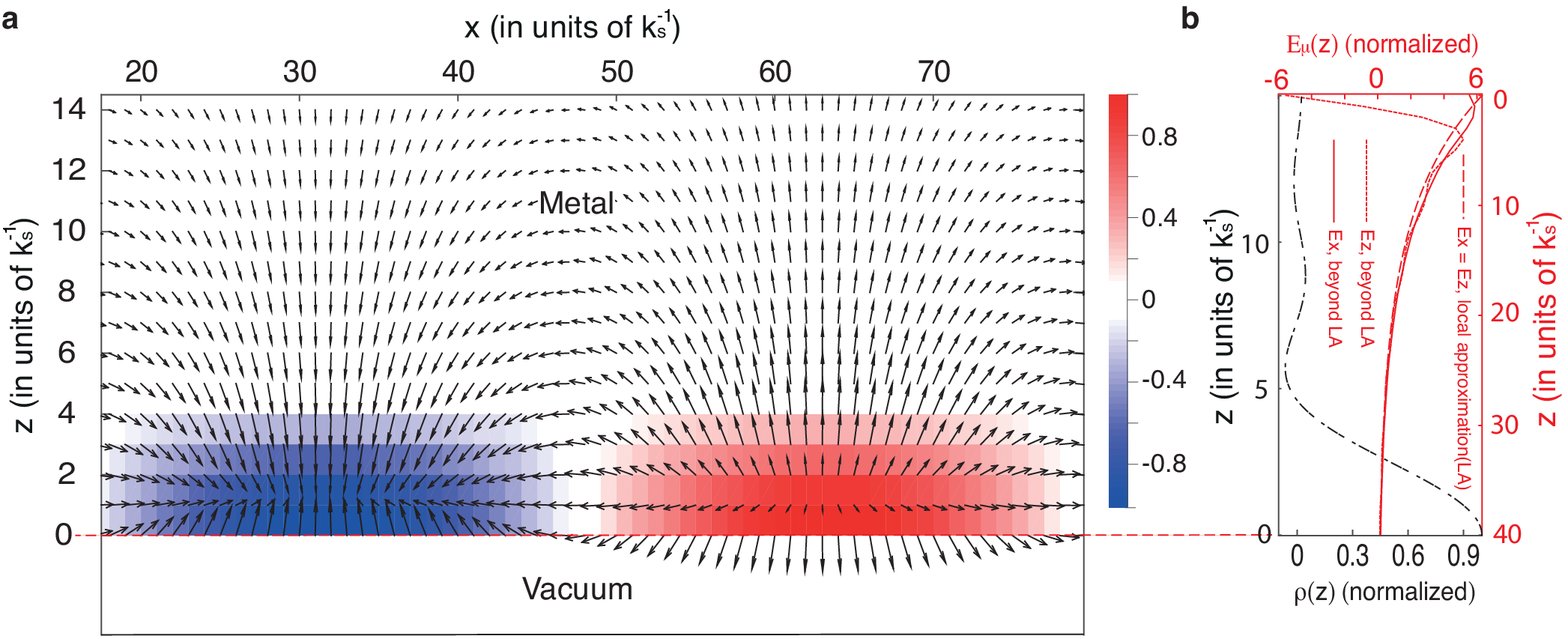}
\end{center}
\caption{\textbf{Properties of surface plasma waves (SPWs) supported on the surface of a metal.} Calculations are presented for $\kappa=k/k_s=0.1$. \textbf{a}, charge density map (color) and electric field map (arrows). The metal, occupying the half space $z\geq0$, interfaces with the vacuum at $z=0$. The charge density $\rho(\mathbf{r},z)$ is calculated via equation (\ref{allequations}) and the electric field $\mathbf{E}(\mathbf{r},z)$ is then obtained by the laws of electrostatics. Though they are strongly concentrated about the surface, the charges do penetrate into the bulk and form a layer that is a few multiples of $k^{-1}_s$ thick, where $k_s=\omega_s/v_F$. \textbf{b}, plots of $\rho(z)$ and $E_{\mu}(z)$, where $\mu=x,z$. $\rho(z)$ has been normalized by its value at the surface while $E_{\mu}(z)$ by the surface charge density $\rho_s$. The actual electric field $\mathbf{E}$ (dotted and solid lines) differs from what would be expected from the local approximation (LA, dashed lines). In particular, $E_z$ (dotted line) changes sign across the surface layer and hence $E_z(0)$ equals $-2\pi\rho_s$ rather than $2\pi\rho_s$. \label{figure:f1}}
\end{figure*}

To vindicate the claim, we consider an ideal yet prototypical system: a semi-infinite metal (SIM) occupying the half space $z\geq 0$ and interfacing with the vacuum at a geometrical surface $z=0$ (Fig.~\ref{figure:f1}). Throughout we use $\mathbf{r}=(x,y)$ for planar coordinates. For simplicity, we assume that inter-band transitions in the metal are suppressed near the SPW frequency $\omega_s$ so that the electrons execute only intra-band motions and can be treated as nearly free. Their kinetic energies are given by $\varepsilon(\mathbf{v})=\frac{1}{2}m\mathbf{v}^2$, where $m$ and $\mathbf{v}$ denote the mass and velocity of the electrons, respectively. The surface is supposed to be perfectly smooth so that all incident electrons are specularly reflected~\cite{reuter1948,ziman1960}. The effects of inter-band transitions, surface roughness and dielectrics in place of the vacuum will be discussed afterwards. If the metal is so clean that the electronic collision rate $\tau^{-1}$ is much smaller than $\omega_s$, the self-amplification rate can be analytically calculated and is given by
\begin{equation}
 \gamma=\frac{\omega_s}{2\pi}\left(A_0-A_1\kappa\tan^{-1}(\kappa^{-1})\right), \label{arate}
\end{equation}
where $A_0 \approx A_1\approx 3$ are constants, $\omega_s$ is the SPW frequency, $k$ is the SPW wave number, $v_F$ is the Fermi velocity of electrons and
$\kappa = k/k_s$ with $k_s=\omega_s/v_F$ of the order of Fermi
wave number $k_F$. This expression is valid for $k_0<k<k_s$, where $k_0=\omega_s/c$ with $c$ being the speed of light. It suggests that a measurement of
$\gamma$ can be used to learn about the Fermi surface of metals,
in close analogy with the anomalous skin effect, whereby the skin depth
furnishes similar information~\cite{kaganov1997}. 

To establish equation (\ref{arate}), which constitutes the key message of the present work, we study the dynamics of the charge density $\rho(\mathbf{r},z,t)$ in the metal rather than that of the electromagnetic field $\mathbf{E}(\mathbf{r},z,t)$ generated by it. As to be seen later, this approach offers a great advantage when it comes to identifying the SPW solutions. Without loss of generality, all quantities are prescribed in the form of plane waves propagating along positive $x$ direction with frequency $\omega$. To ease the notation, we write $\rho(\mathbf{r},z,t) = \rho(z)~e^{i(kx-\omega t)}$ and similarly for all other quantities. In the regime of linear response, the electrical current density is given by
\begin{equation}
J_{\mu}(z) = \Theta(z)~\int^{\infty}_0dz'~\sigma_{\mu}(z,z';\omega)~E_{\mu}(z'), 
\label{1}
\end{equation}  
where $\sigma_{\mu}$, $J_{\mu}$ and $E_{\mu}$ are the $\mu$-th component
of the electrical conductivity, $\mathbf{J}$ and $\mathbf{E}$, respectively. We have
included in equation (\ref{1}) the Heaviside step function $\Theta(z)$ to ensure
that no current flows beyond the surface. This constraint is
automatically satisfied if $\sigma_{\mu}(z,z';\omega)$ identically vanishes on the surface, 
which would exclude the existence of SPWs.

In existing discourse of SPWs, a local approximation for the electrical conductivity has been unanimously adopted~\cite{pitarke2007}. In order to illustrate our approach, we first discuss this approximation, by which we write
$\sigma_{\mu}(z,z';\omega)\approx \sigma_D(\omega) \delta(z-z')$, with
$\delta(z-z')$ being the Dirac function and $\sigma_D(\omega) =
(\omega^2_p/4\pi)(i/\tilde{\omega})$ the Drude
conductivity, where $\tilde{\omega}=\omega+i/\tau$ and
$\omega_p=\sqrt{4\pi n_0e^2/m}$ is the bulk plasma wave
frequency. We have denoted by $n_0$ the mean density and
by $e$ the charge of the electrons. 
In the bulk, this approximation may be justified if the SPW wavelength $2\pi/k$ is much longer than the electronic mean free path $l=v_F\tau$, so that during an oscillation the electrons have suffered sufficient collisions and they produce only a diffusive current. Near the surface, however, this approximation has to be abandoned, because the surface breaks the translational symmetry and there is always a layer of electrons emerging from beneath and ballistically moving away. With this approximation, equation (\ref{1}) becomes $J_{\mu}(z) \approx
\sigma_D(\omega)\Theta(z)E_{\mu}(z)$. Applying the continuity equation and Maxwell's equations, we find the equation of motion for the charge density as follows
\begin{equation}
\left(\tilde{\omega}^2 - \omega^2_p\right) \rho(z)=
 -i\tilde{\omega}J_z(0)\delta(z), 
\label{2}
\end{equation}  
where $\delta(z)$ arises from an abrupt change in the conductivity across the surface. Solutions to equation (\ref{2}) fall in two categories, depending on whether $J_z(0)$ vanishes or not. The solutions with vanishing $J_z(0)$ are no more than standing bulk plasma waves, whose eigenfrequencies are $\omega_p$ in the collisionless limit where $\tau\rightarrow\infty$. The rest of the solutions have $\rho(z) = \rho_s\delta(z)$ totally localized at the surface and they represent SPWs. Here  
\begin{equation}
\rho_s =
 i\tilde{\omega}J_z(0)\left(\omega^2_p-\tilde{\omega}^2\right)^{-1} \label{3}
 \end{equation}
gives the surface charge density. The celebrated SPW eigenfrequency $\omega_s=\omega_p/\sqrt{2}$ is obtained from equation (\ref{2}) in the electrostatic limit where $k>k_0$ and in the collisionless limit~\cite{note}. The electric field exponentially decays into the bulk with a length $k^{-1}$.

SPWs are dynamical modes of a compounded system consisting of both the electrons and the electric field. Two general approaches can then be envisaged to study them: one may either eliminate the electrons in favor of the field or do the converse. In the conventional approach, the electrons are eliminated yielding an equation of only the field. The SPWs are embodied in an \textit{ansatz}, which assumes that the corresponding solutions have an exponentially decaying field on each side of the surface and the two sides are then adjoined by boundary conditions~\cite{raether1988,maier2007,pitarke2007}. This \textit{ansatz}, unfortunately, is valid only in the local approximation, beyond which the field can not be written as an exponential function in the metal any more (Fig.~\ref{figure:f1}). This difficulty in identifying SPWs is avoided in the present approach, whereby the field is eliminated. The resulting equation for the charge density has such a structure that SPWs are always recognized as localized solutions satisfying $J_z(0)\neq 0$, and these solutions are unequivocally separated from the extended solutions that stand for bulk waves. This feature reminds the readers of Schr\"{o}dinger's equation in quantum mechanics describing a particle moving in a point-like potential field, wherein localized quantum states are recognized in a similar fashion. It spells a great advantage of the present method. 

Although the local approximation has been unanimously adopted in studying SPWs, it fails to capture the ballistic electronic motions, which differ from diffusive motions in a qualitative manner. Below we show that ballistic currents cause an instability of SPWs and therefore provide a self-amplification channel. We employ Boltzmann's equation to evaluate the conductivity so that ballistic motions are taken into account~\cite{ziman1960}. We denote by $f(\mathbf{v},\mathbf{r},z,t)=f_0\left(\varepsilon(\mathbf{v})\right) + g(\mathbf{v},z)e^{i(kx-\omega t)}$ the electronic distribution function, where  $f_0(\varepsilon)$ is the equilibrium Fermi-Dirac function and $g(\mathbf{v},z)$ represents the non-equilibrium part. To the linear order in $\mathbf{E}(z)$ and within the relaxation time approximation, Boltzmann's equation can be written as
\begin{equation}
\frac{\partial g(\mathbf{v},z)}{\partial z} + \lambda^{-1}~g(\mathbf{v},z) +
 e~\frac{\mathbf{v}\cdot\mathbf{E}(z)}{v_z}~\frac{\partial f_0}{\partial
 \varepsilon(\mathbf{v})} = 0,
\label{4}
\end{equation}
where $\lambda =
\frac{v_z\tau}{1-i\omega\tau}$ for small $k$. Together with the boundary conditions suitable for smooth surfaces~\cite{reuter1948,ziman1960}, equation (\ref{4}) can be solved in a standard
manner to obtain $g(\mathbf{v},z)$, from which the current density $J_{\mu}(z) = e(m/2\pi\hbar)^3~\Theta(z)~\int
d \mathbf{v}~v_{\mu}~g(\mathbf{v},z)$ and then the conductivity 
$\sigma_{\mu}$ can be computed~\cite{reuter1948}. At zero temperature we obtain
$\sigma_{\mu}(z,z';\omega) = \frac{3\omega^2_p}{4\pi v_F}
 \left(\Sigma_{\mu}(z+z';\Lambda) + \Sigma_{\mu}(z-z';\Lambda)\right)$, 
where $\Lambda = \frac{v_F\tau}{1-i\omega\tau}$,
$\Sigma_x(z;\Lambda) = \Sigma_y(z;\Lambda) = \mathcal{E}_1(|z|/\Lambda)
- \mathcal{E}_3(|z|/\Lambda)$ and $\Sigma_z(z;\Lambda) =
2\mathcal{E}_3(|z|/\Lambda)$ with $\mathcal{E}_n(z)$ being the $n$-th
order exponential integral~\cite{abramowitz1965}. 

\begin{figure*}
\begin{center}
\includegraphics[width=0.9\textwidth]{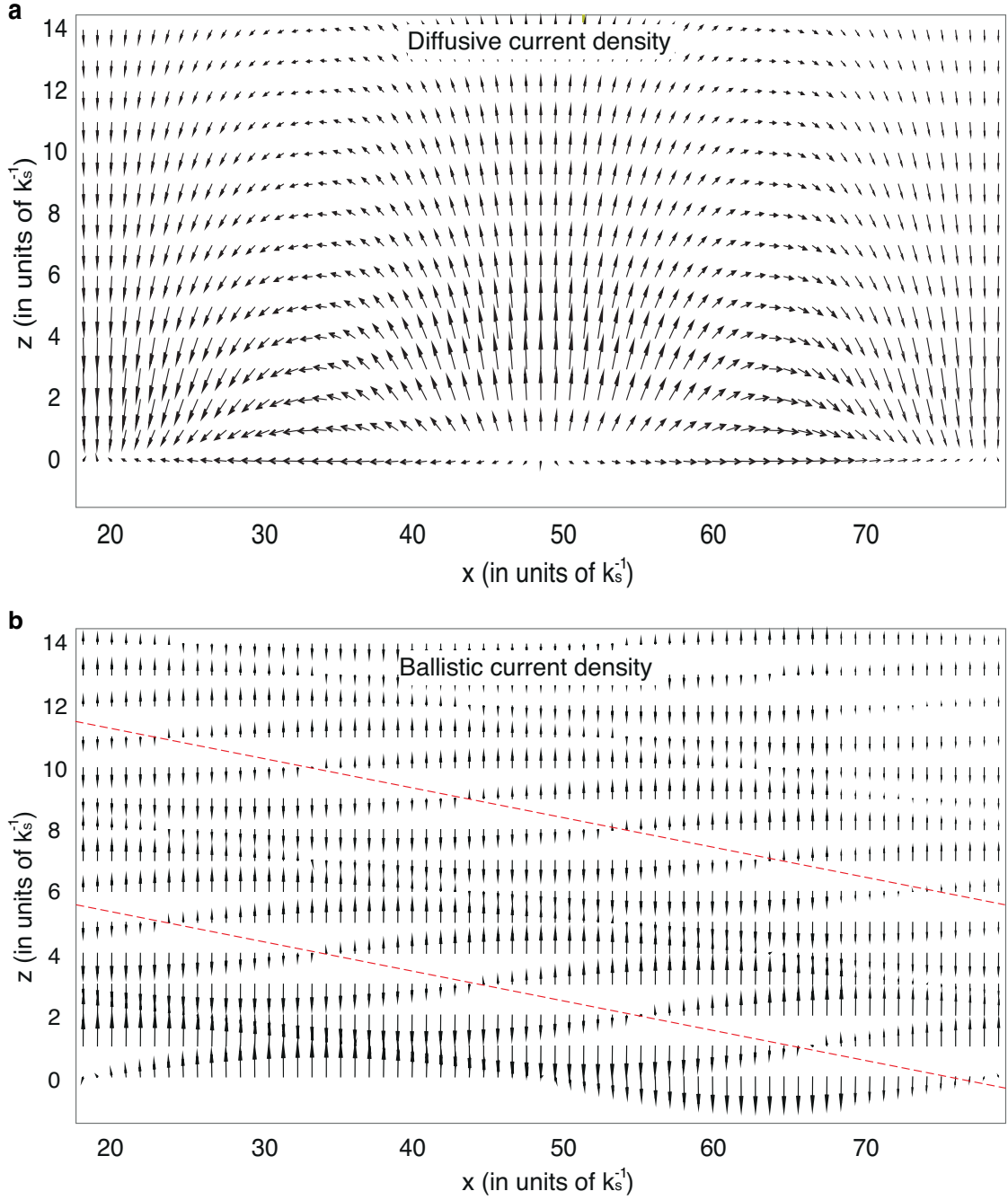}
\end{center}
\caption{\textbf{Diffusive versus ballistic currents in collisionless SPWs.} $\kappa=0.1$ is used. \textbf{a}, diffusive current density map. By comparison with Fig.~\ref{figure:f1}, one sees that the diffusive current $\mathbf{J}_D(z)$ always differs from the local electric field $\mathbf{E}(z)$ in phase by $\pi/2$ and then they are locally perpendicular to each other. There is no dispersion along $z$. Within the surface layer, the $z$-component of $\mathbf{J}_D(z)$ changes sign, without which SPWs would not exist. This is in accord with the fact that $E_z(z)$ changes sign. \textbf{b}, ballistic current density map. The ballistic current $\mathbf{J}_B(z)$ almost always points normal to the surface. As indicated by the dashed lines, it disperses along $z$-axis. This is because $\mathbf{J}_B(z)$ has an extra phase $\phi(z)=\omega z/v_F$. \label{figure:f2}}
\end{figure*}

The current density $\mathbf{J}(z)$ naturally splits in two portions: one is the diffusive current $\mathbf{J}_{D}(z)$ while the other is the ballistic current $\mathbf{J}_B(z)$. What distinguishes them is their distinctly different phase behaviors. $\mathbf{J}_D(z)$ has a phase that is always at odds with $\mathbf{E}(z)$ by only $\pi/2$ in the collisionless limit. By contrast, $\mathbf{J}_B(z)$ disperses along $z$ and possesses an extra phase $\phi(z)=\omega z/v_F$. Obviously, this phase is accumulated during the journey of an electron which travels ballistically, i.e. without suffering any collisions, from the surface to a depth $z$. With this criteria an exact formulation of $\mathbf{J}_{D}$ and $\mathbf{J}_B$ can be established (see Supplementary Information). Numerical calculations of these currents have been conducted and the result is exhibited in Fig.~\ref{figure:f2}. By comparison with the electric field map shown in Fig.~\ref{figure:f1} \textbf{a}, the as-described phase behaviors become clear. Now that $\mathbf{J}_D(z)$ is always perpendicular to $\mathbf{E}(z)$, no net work can be done on the electrons by $\mathbf{E}$ via $\mathbf{J}_D$. Nevertheless, this is not so with $\mathbf{J}_B(z)$ due to the extra phase $\phi(z)$. As shown in Supplementary Information, $\mathbf{J}_B(z)$ allows the electric field to extract a positive amount of energy from the electrons. As such, $\mathbf{E}(z)$ must grow in time and this signifies an instability of the system. In what follows, we discuss the instability using the equation of motion for $\rho(z)$. We work out the eigenfrequency of the system and show that it has a positive imaginary part $\gamma$~\cite{note2}.

To demonstrate the instability, it suffices to neglect the spatial penetration of $\rho(z)$ and take $\rho(z) \approx \rho_s\delta(z)$. Substituting $E_z(z) \approx -2\pi\rho_se^{-kz}$ in equation (\ref{1}) and using the expression of $\sigma_{\mu}$ given above, we obtain $J_z(0)\approx -2\pi\rho_s\sigma_D(\omega)
\left(1-\frac{3}{4}k\Lambda\right)$. Plugging this in equation (\ref{3}), we find $\omega\approx \omega_s+i~\frac{3}{8}kv_F$ in the collisionless limit. This shows that SPW can be unstable. The resulting amplification rate, however, is only an underestimate. 

A more accurate and much bigger amplification rate can be obtained by taking care of the spatial penetration of $\rho(z)$. We set out to find the exact equation of motion for $\rho(z)$. This can be achieved again using the equation of continuity and the laws of electrostatics. As shown in Supplementary Information, the desired equation has the same structure as equation (\ref{2}), apart from a generalization of $\omega^2_p$ to a non-local form: $\omega^2_p \rho(z) \rightarrow\omega^2_p\int dz'~w(z,z')\rho(z')$, where $w(z,z')$ is a kernal that depends on $\Lambda$. The exact expression of $w(z,z')$ can be found
in Supplementary Information. Defining $\rho(-z) = \rho(z)$, we can write
$\rho(z) = \frac{2}{\pi} \int^{\infty}_0dq~\rho_q\cos(qz)$ with $\rho_q$
being the Fourier component of $\rho(z)$. The
equation for $\rho_q$ can be established by straightforward
manipulations and we have
\begin{subequations}
\label{allequations}
\begin{eqnarray}
&~&\left(\tilde{\omega}^2-\omega^2_pW(q\Lambda)\right)\rho_q =
 -i\tilde{\omega}J_z(0),\label{equationa}\\&~& ~~
 J_z(0)=i~\frac{\omega^2_p}{\tilde{\omega}}\int^{\infty}_0\frac{dq'}{\pi}\frac{k}{k^2+q^{'2}}~G(q',k)~\rho_{q'},\label{equationb}
\end{eqnarray}
\end{subequations}
where $W(q\Lambda)=\int^{\infty}_1\frac{dt}{t^2}\frac{3}{t^2+q^2\Lambda^2}$
arises from a double Fourier transform of $w(z,z')$ and
\begin{equation}
G(q,k) =
 \int^{\infty}_1dt~\frac{3}{t^3}\left(\frac{2(q/k)^2(k\Lambda)}{t^2+(q/k)^2(k\Lambda)^2}-\frac{1}{t+k\Lambda}\right). \label{8}
\end{equation}
The local approximation can be restored by taking $\tau=0$ (i.e. $\Lambda=0$). In the collisionless limit, $\Lambda\rightarrow i(v_F/\omega)$ becomes imaginary and $W(q\Lambda)$ becomes real. The eigenfrequency of bulk plasma waves is determined by $\omega^2 = \omega^2_pW(q\Lambda)$ and therefore also becomes real. As expected, ballistic motions do not damp or amplify bulk waves.

\begin{figure*}
\begin{center}
\includegraphics[width=0.9\textwidth]{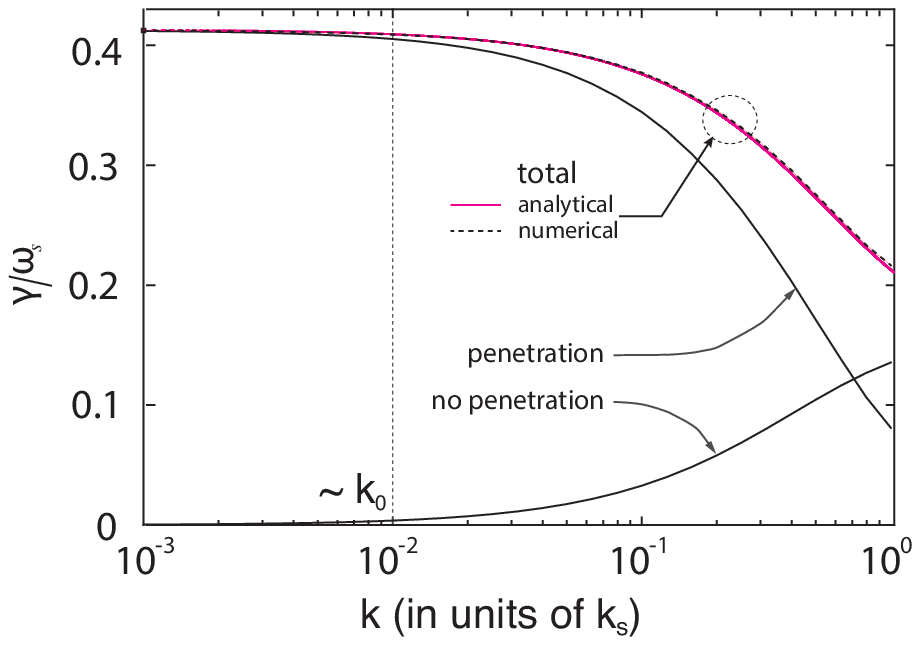}
\end{center}
\caption{\textbf{Amplification rate $\gamma$ for SPWs neglecting losses.} $\gamma$ would be significantly underestimated were the charge penetration effects ignored. The analytical expression (solid curve) as given by equation (\ref{arate}) well agrees with numerical calculations (dotted curve). Here $k_0=\omega_s/c$, with $c$ being the speed of light in vacuum. Typically, $k_0/k_s=v_F/c \sim 0.01$. \label{figure:f3}}
\end{figure*}

SPWs are represented by the localized solutions of equation (\ref{allequations}). Expressing $\rho_q$ in terms of $J_z(0)$ via equation (\ref{equationa}) and inserting
it in equation (\ref{equationb}), we find
\begin{equation}
1 =
 \int^{\infty}_0\frac{dq'}{\pi}\frac{k}{k^2+q^{'2}}\frac{\omega^2_pG(q',k)}{\tilde{\omega}^2-\omega^2_pW(q'\Lambda)},
\label{9}
\end{equation}   
which can be regarded as the secular equation for SPWs. In the collisionless limit, $\gamma$ can only come from $G^{''}$, the
imaginary part of $G = G'+iG^{''}$. See that
$G'(q,k)=-\int^{\infty}_1\frac{dt}{t^2}\frac{3}{t^2+(kv_F/\omega)^2}\approx
-1$. If $G^{''}$ were dropped and then $W(q'\Lambda)\approx 1$, we would immediately revisit from equation (\ref{9}) the
expression of the SPW frequency $\omega_s$. For
$\omega$ in the vicinity of $\omega_s$, $G^{''}$ is a small quantity with two parts:
$G^{''}(q,k)\approx(kv_F/\omega)\left(\frac{3}{4}+\int^{\infty}_1\frac{dt}{t^3}\frac{6(q/k)^2}{t^2+(qv_F/\omega)^2}\right)$. The
first part was already obtained when the spatial penetration of $\rho(z)$ had been neglected, while the second part arises from penetration effects. The solution to equation (\ref{9}) can be found if $\gamma$ is smaller than $\omega_s$. We find $\gamma=
F\left(\frac{kv_F}{\omega_s}\right)~kv_F$ with $F(\kappa)$ being an
integral. Performing the integral, we immediately arrive at equation (\ref{arate}), thereby proving our main statement. A plot of $\gamma$ is displayed in Fig.~\ref{figure:f3}. We see that penetration effects considerably increase the magnitude of $\gamma$.

The theory can also be applied to rough
surfaces. Electrons emerging from such surfaces bear no
memory of their previous trajectories and the conductivity
only consists of the second term of $\sigma_{\mu}$. The equation of motion
for $\rho_q$ can be shown of the same form as
equation (\ref{allequations}) but with a
slightly altered $G(q,k)$. The real part of $G$ is almost unchanged but the imaginary part just gets halved, implying that $\gamma$ is also halved. 

If we replace the vacuum by a dielectric $\epsilon$, $\gamma$ will
be reduced by an order of ~$\epsilon^{-1}$ due to weakened $E_z(z)$ and $J_z(0)$. Roughly speaking, $\rho(z)$ present on the metal side
induces mirror charges amounting to
$\rho'(z)=-\frac{\epsilon-1}{\epsilon+1}\rho(-z)$ on the dielectric
side. If $\rho(z)$ is highly localized about the interface, as is
with SPWs, $\rho(z)$
and $\rho'(z)$ will combine to give a net charge of
$\frac{2}{\epsilon+1}\rho(z)$. As a result, $E_z$ and $J_z(0)$ will be
reduced by a factor of $\frac{2}{\epsilon+1}$. This leads to smaller
$\gamma$ and smaller $\omega_s=\omega_p/\sqrt{\epsilon+1}$. For SPWs, $\rho(z)$ can be calculated by equation 
(\ref{equationa}). We display $\rho(z)$ in Fig.~\ref{figure:f1} \textbf{b}, where we see that $\rho(z)$ is
localized within a few multiples of $k^{-1}$ of the
surface. The profile of $\rho(z)$ does not depend much on the
dielectric, because it is basically determined by the bulk waves via the
factor $W(iq/k_s)$.   

We may generalize the theory to account for the effects of inter-band transitions by including in the conductivity a contribution from inter-band transitions. Going through the same procedures, one can show that the structure of the equation of motion for $\rho_q$ [i.e. equation (\ref{allequations})] survives intact. In particular, the bulk waves and SPWs are still identified by the criteria whether $J_z(0)$ vanishes or not. The exact forms of $W$ and $J_z(0)$ cannot be the same. Since it describes the properties of bulk waves, $W$ can be obtained in the usual way. Generally, $W$ becomes complex and bulk waves are damped. The imaginary part of $W$ can readily be absorbed in the definition of $\tau$. $J_z(0)$ can be split in two parts, one from intra-band motions [given by equation (\ref{equationb})] while the other from inter-band transitions. Calculating the latter is a formidable task, which shall be pursued in the future. Despite this, inter-band transitions do not destroy the ballistic currents and the amplification channel. 

We have thus shown that SPWs can possibly amplify themselves by drawing energy from the bulk. The mechanism originates with ballistic currents. It is generic and robust. We expect similar phenomena to take place in other systems such as metallic films and nanoparticles as well as atomically thin layers like graphene. Our work not only discloses a fundamental property of SPWs but also can have numerous practical implications to be explored in the future. 

\noindent
\textbf{Acknowledgement} \\ HYD acknowledges the International Research
Fellowship of the Japan Society for the Promotion of Science
(JSPS). This work is supported by JSPS KAKENHI Grant No. 15K13507 and
MEXT KAKENHI Grant No. 25107005.

\noindent
\textbf{Author Contributions} \\ HYD conceived and performed the study. He also wrote the manuscript. All authors participated in discussions of the work and in finalizing the manuscript.

\newpage

\newpage

\newpage


\begin{thebibliography}{9}
\bibitem{note2}
Of course $\gamma$ can also be calculated directly from the rate of energy conversion. Details along this line will be published elsewhere.
\end{thebibliography}
\end{document}